\documentclass[print]{revtex4}
\usepackage{amsmath,amssymb}
\usepackage{graphicx}

\draft

\begin{document}

\title{\large \bf Possibility of Inhomogeneous Coupling Leading to Decoherence in
an Electromagnetically-Induced-Transparency Quantum-Memory
Process}

\author{Xiong-Jun Liu$^{a,b}$\footnote{Electronic address: phylx@nus.edu.sg}, Zheng-Xin
Liu$^{b,c}$, Xin Liu$^{b,c}$ and Mo-Lin Ge$^{b,c}$}

\affiliation{ a. Department of Physics, National University of
Singapore, 2 Science Drive 3, Singapore 117542 \\
b.Theoretical Physics Division, Nankai Institute of
Mathematics,Nankai University, Tianjin 300071, P.R.China\\
c. Liuhui Center for Applied Mathematics, Nankai University and
Tianjin University, Tianjin 300071, P.R.China}

\begin{abstract}
The effect of inhomogeneous coupling between three-level atoms and
external light fields is studied in the electromagnetically
induced transparency (EIT) quantum memory techqnique. By
introducing a subensemble-atomic system to deal with present
inhomogeneous coupling case, we find there is a non-symmetric
dark-state subspace (DSS) that allows the EIT quantum memory
technique to function perfectly. This shows that such memory
scheme can work ideally even if the atomic
state is very far from being a symmetric one.\\

\end{abstract}

\baselineskip=16pt

\maketitle

\indent

\section{introduction}

Recently, the novel mechanism of electromagnetically induced
transparency (EIT) \cite{2} and its many important applications
have attracted much attention in both experimental and theoretical
aspects \cite{3,4,5,wu}. In particular, based on the ``dark-state
polaritons" (DSPs) theory \cite{6}, the quantum memory via EIT
technique is actively being explored by transferring the quantum
states of photon wave-packet to metastable collective
atomic-coherence (collective quasi spin states) in a loss-free and
reversible manner \cite{7}. Quantum information processing based
on the interaction between light fields and large number of atoms
indicates a large dynamical symmetry \cite{Lidar}. For the
three-level EIT quantum memory technique, a semidirect product
group under the condition of large atom number and low collective
excitation limit \cite{6} is discovered by Sun $et al.$ \cite{8},
and the validity of adiabatic condition for the evolution of DSPs
is also confirmed. Subsequently, a series of works are involved in
the hidden dynamical symmetry and the applications to quantum
information processing, such as the generation of quantum
entanglement between atoms (or lights) with multi-level atomic
system (or multi-atomic-ensemble system) \cite{liu,atom,liu1},
etc.

However, to realize such a large dynamicl symmetry in the
field-atoms interaction model, the coupling between the external
fields and atomic ensemble is usually assumed to be homogeneous in
DSP theory, i.e. the effect of inhomogeneity for atoms with
different spatial positions is ignored. Very recently, Sun $et
al.$ \cite{sun} investigated the decoherence for a superposition
of symmetric internal states of a two-level atomic gas due to the
inhomogeneous coupling with external light fields. Within their
model, for an ensemble composed of $N$ atoms, they find that the
apparent decoherence or dissipation rate for superpositions of
collective spin states scales as $\sqrt{N}$. On the other hand,
the applications to magnetometery with inhomogeneous coupling
between the two-level atomic ensembles and light field was studied
by A. Kuzmich et al \cite{kuzmich}. Also, the effect of
inhomogeneous coupling in spin squeezing and precision measurement
with light was studied by L. B. Madsen and K. M{\o}lmer
\cite{madsen}. As a result, the effect of inhomogeneous coupling
in EIT quantum memory process naturally becomes an important
issue, especially about the validity in this memory process with
collective spin states \cite{sun}, which we proceed to study in
the following.

In this paper, we show the model for three-level EIT quantum
memory with inhomogeneous coupling can be equivalently treated as
a many-atomic-ensemble system (sub-ensemble atomic system), and
then a non-symmetric dark-state subspace (DSS) of present system
can be exactly obtained. Although the leakage coefficient will be
larger than zero after the quantized probe light is stored, the
quantum states of the probe light can be fully recovered when it
is released by turning on the control field again.

\section{model}
We consider a quasi one-dimensional model for an ensemble of $N$
identical atoms with the three-level $\Lambda$ type structure. The
$j$-th atom interacts with an input quantized field with coupling
constant $g_j$, and the classical control filed with
time-space-dependent Rabi-frequency $\Omega(z_j,t)$. The
interaction hamiltonian is
\begin{eqnarray}\label{eqn:h1}
\hat H=-\sum^N_jg_j\Bigl(\hat\sigma_{ab}^j\, \hat
E^{(+)}(z_j,t)+h.a\Bigr) -\hbar \sum^N_j\Bigl(\hat\sigma_{ac}^j\,
\Omega(z_j,t) \, {\rm e}^{i (k_cz_j-\nu_c t)}+h.a\Bigr)
\end{eqnarray}
where $g_j$ and $\Omega(z_j)$ are the inhomogeneous coupling
constants between the $j$-th atom and probe and control fields.
Although here we consider the inhomogeneous coupling case,
generally the coupling constants should be slowly varying with the
adjacent atoms. For this we make an approximation that in a small
length $\Delta z_{\sigma}$ around the point $z_{\sigma}$ there are
$N_{\sigma}\gg1$ atoms which interact with the quantized field
with a homogeneous coupling constant $g_{\sigma}$, and the
classical control filed with time-dependent Rabi-frequency
$\Omega(z_{\sigma},t)$. Therefore, the present model with
inhomogeneous coupling can be equivalently described as the
many-atomic-ensemble case (see the ref. \cite{liu1} section II,
part B), i.e. all the atoms in the length $\Delta z_{\sigma}$
(with atom number $N_{\sigma}$) can be approximately described as
a single atomic ensemble with the same coupling constant
$g_{\sigma}$ and Rabi-frequency
$\Omega_{\sigma}(t)=\Omega(z_{\sigma},t)$ (while the couplings are
different between different ensembles)
\begin{eqnarray}\label{eqn:tranformation1}
\sum_{z_j\in N_{\sigma}}g_j\hat\sigma_{\mu\nu}(z_j)\rightarrow
g_{\sigma}\sum_{z_j\in N_{\sigma}}\hat\sigma_{\mu\nu}(z_j),
\end{eqnarray}
\begin{eqnarray}\label{eqn:tranformation2}
\sum_{z_j\in
N_{\sigma}}\Omega(z_j,t)\hat\sigma_{\mu\nu}(z_j)\rightarrow
\Omega_{\sigma}(t)\sum_{z_j\in
N_{\sigma}}\hat\sigma_{\mu\nu}(z_j).
\end{eqnarray}
This assumption is reasonable. Because, for example, we can
consider the case that the atoms in a very small volume $\Delta V$
around $z_{\sigma}$ point (the number is still larger larger than
1, see e.g. \cite{6,7}) have the nearly equal coupling constants,
so that the coupling in the small volume can be treated as
homogeneous (we may call this the quasi-inhomogeneous case). Then,
considering all transitions at resonance and for the single-mode
probe field case, i.e. ${\hat E}^{(+)}(z_j,t)=\hat a
e^{i(k_pz_j-\nu_pt)}$, the interaction Hamiltonian can be
rewritten as:
\begin{eqnarray}\label{eqn:1}
\hat H=\sum_{\sigma=1}^mg_{\sigma}\sqrt{N_{\sigma}}\hat a\hat
A_{\sigma}^{\dag}+\sum_{\sigma=1}^m\Omega_{\sigma}(t)\hat
T^+_{\sigma}+h.c.,
\end{eqnarray}
where $N_1+N_2+...+N_m=N$ and the subscript $\sigma=1,2,3,...,m$
denotes the corresponding atomic ensemble and the collective
atomic excitation operators:
\begin{equation}\label{eqn:2}
\hat A_{\sigma}=\frac{1}{\sqrt{N_{\sigma}}}\sum_{z_j\in
N_{\sigma}}e^{-i(k_{ba}z_j-\omega_{ba}t)}\hat\sigma_{ba}^{j(\sigma)},
\ \ \hat C_{\sigma}=\frac{1}{\sqrt{N_{\sigma}}}\sum_{z_j\in
N_{\sigma}}e^{-i
(k_{bc}z_j-\omega_{bc}t)}\hat\sigma_{bc}^{j(\sigma)}
\end{equation}
with $\hat\sigma^i_{\mu\nu}=|\mu\rangle_{ii}\langle\nu|
(\mu,\nu=a,b,c)$ being the flip operators of the $i$-th atom
between states $|\mu\rangle$ and $|\nu\rangle$, $\bf k_{ba}$ and
$\bf k_{ca}$ are, respectively, equal the wave vectors of the
quantum and classical light fields, $\bf k_{bc}=\bf k_{ba}-\bf
k_{ca}$ and
\begin{equation}\label{eqn:3}
\hat T^{-}_{\sigma}=(\hat T_{\sigma}^{+})^{\dagger}=\sum_{z_j\in
N_{\sigma}}e^{-i(k_{ca}z_j-\omega_{ca}t)}\hat\sigma
_{ca}^{j(\sigma)}.
\end{equation}
Denoting by
$|b^{({\sigma})}\rangle=|b^{({\sigma})}_1,b^{({\sigma})}_2,...,b^{({\sigma})}_{N_{\sigma}}\rangle
({\sigma}=1,2,...,m)$ the collective ground state of the
${\sigma}$-th atomic ensemble with all atoms staying in the same
single particle ground state $|b\rangle$, we can easily give other
quasi-spin wave states by the operators defined in formula
(\ref{eqn:2}): $|a^n_{(\sigma)}\rangle=[n!]^{-1/2}(\hat
A_{\sigma}^{\dag})^n|b^{(\sigma)}\rangle$ and
$|c^n_{(\sigma)}\rangle=[n!]^{-1/2}(\hat
C_{\sigma}^{\dag})^n|b^{(\sigma)}\rangle$. Similarly, in large
$N_{\sigma}$ limit and low excitation condition, it follows that
$[\hat A_{(i)},\hat A^{\dag}_{(j)}]=\delta_{ij}, [\hat
C_{(i)},\hat C^{\dag}_{(j)}]=\delta_{ij}$ and all the other
commutators are zero, which shows the mutual independence between
these bosonic operators $\hat A_{i}$ and $\hat C_{i}$. On the
other hand, one can easily find the commutation relations: $[\hat
T^+_{i},\hat T^-_{j}]=\delta_{ij}\hat T^z_{j}$ and $[\hat
T^z_{i},\hat T^{\pm}_{j}]=\pm\delta_{ij}\hat T^{\pm}_{j}$, where
\begin{equation}\label{eqn:4}
\hat T_{\sigma}^{z}=\sum_{z_j\in N_{\sigma}}(\hat\sigma
_{aa}^{j(\sigma)}-\hat\sigma _{cc}^{j(\sigma)})/2, \
(\sigma=1,2,...,m)
\end{equation}
are the traceless operators.

Following the results obtained in refs. \cite{liu1}, the
non-symmetric DSP operator of present system can be defined as
\begin{equation}\label{eqn:d1}
\hat d=\cos\theta\hat a-\sin\theta\prod_{j=1}^{m-1}\cos\phi_j\hat
C_1-\sin\theta\sum_{k=2}^m\sin\phi_{k-1}\prod_{j=k}^{m-1}\cos\phi_j\hat
C_k,
\end{equation}
where the mixing angles $\theta$ and $\phi_j$ are defined through
\begin{equation}\label{eqn:5}
\tan\theta=\frac{\bigr[\sum_{j=1}^{m}\bigr(g_j^2N_j\prod_{k=1,k\neq
j}^{m}\Omega^2_k\bigr)\bigr]^{1/2}}{\Omega_1\Omega_2...\Omega_m}
\end{equation}
and
\begin{equation}\label{eqn:6}
\tan\phi_j=\frac{g_{j+1}\sqrt{N_{j+1}}\prod_{k=1}^{j}\Omega_k}{\bigr[\sum_{k=1}^{j}\bigr(g^2_kN_k\prod_{s=1,s\neq
k}^{j+1}\Omega^2_s\bigr]^{1/2}}
\end{equation}
From the equation (\ref{eqn:6}) one finds
$\tan\phi_1=g_2\sqrt{N_2}\Omega_1/g_1\sqrt{N_1}\Omega_2, \
\tan\phi_2=g_3\sqrt{N_3}\Omega_1\Omega_2/\sqrt{g_1^2N_1
\Omega_2^2\Omega_3^2+g_2^2N_2\Omega_1^2\Omega_3^2} ...$, etc.
Also, by a straightforward calculation one can verify that $[\hat
d,\hat d^{\dag}]=1$ and $ [\hat H,\hat d \ ]=0 $, hence the
general atomic dark states can be obtained through
$|D_n\rangle=[n!]^{-1/2}(\hat d^{\dag})^n|b^{(1)}, b^{(2)},...,
b^{(m)}\rangle_{atom}\otimes|0\rangle_{photon}$, where the
collective ground state
$|b^{(\sigma)}\rangle=|b_1,b_2,...,b_{N_{\sigma}}\rangle$ and
$|0\rangle_{photon}$ denotes the electromagnetic vacuum of the
quantized probe field. With the exact dark state obtained in
present inhomogeneous EIT model, we can readily study the effect
of inhomogeneous coupling between atoms and external fields.

\section{The effect of the inhomogeneous coupling}

To verify the effect on the quantum memory processing due to the
inhomogeneous coupling constant, we rewrite the Hamiltonian of the
formular (\ref{eqn:1}) as $H=H_0+H_1$ with
\begin{eqnarray}\label{eqn:hamiltonian2}
\hat H_0=g_0\sqrt{N}\hat a\hat
A^{\dag}+\sum_{\sigma=1}^m\Omega_0(t)\hat T^+_{\sigma}+h.c.,
\end{eqnarray}
\begin{eqnarray}\label{eqn:hamiltonian3}
\hat H_1=\sum_{\sigma=1}^m\delta_{\sigma}\sqrt{N_{\sigma}}\hat
a\hat A_{\sigma}^{\dag}+\sum_{\sigma=1}^m\lambda_{\sigma}\hat
T^+_{\sigma}+h.c.,
\end{eqnarray}
where $\hat A=\sum^m_{\sigma=1}\sqrt{N_{\sigma}}\hat
A_{\sigma}/\sqrt{N}$, $\hat
C=\sum^m_{\sigma=1}\sqrt{N_{\sigma}}\hat C_{\sigma}/\sqrt{N}$,
$g_{\sigma}=g_0+\delta_{\sigma}$ and
$\Omega_j=\Omega_0+\lambda_{\sigma}$. $\delta_{\sigma}$ and
$\lambda_{\sigma}$ denote the inhomogeneous part of the coupling
of the probe and control fields, respectively. For the usual
three-level EIT technique, the coupling is assumed homogeneous and
then $H_1=0$. Noting that
\begin{eqnarray}\label{eqn:U1}
\hat U_0(t)=\exp{(-i\hat H_0t)}, \ \ \hat U(t)=\exp{(-i\hat Ht)}
\end{eqnarray}
are the corresponding time evolution operators. For some initial
state $|\psi(0)\rangle$, the operator $\hat U_0$ leads to the
evolution $|\psi(t)\rangle_0=\hat U_0(t)|\psi(0)\rangle$, while
the actual evolution reads $|\psi(t)\rangle=\hat
U(t)|\psi(0)\rangle$.

According to the treatment of ref. \cite{sun}, the leakage of the
quantum information is defined as
\begin{eqnarray}\label{eqn:leak1}
\xi=1-|\langle\psi(t)|\psi_0(t)\rangle|^2.
\end{eqnarray}
$\xi\rightarrow0$ means no leakage, while $\xi\rightarrow1$
indicates a complete loss of the system coherence and population.
For a two-level atomic ensemble system \cite{sun}, the leakage or
decoherence of the collective spin states is dependent on the
inhomogeneous coupling and number of the atoms. Then there is no
advantage of using collective spin states for quantum information
processing in that system. However, as the discussions in the
following, we will show this is different from the quantum memory
processing via EIT technique  which is concerned with a
three-level atomic ensemble system.

Considering the case of the quantum memory for photons with
present EIT technique, the initial total state reads (meanwhile
$\theta=0$ or the external control field is very strong):
\begin{eqnarray}\label{eqn:ds3}
|\Psi(0)\rangle=\sum_{n}C_n|n\rangle_{photon}\otimes|b^{(1)},
b^{(2)},...,b^{(m)}\rangle_{atom},
\end{eqnarray}
where $|\phi_0\rangle=\sum_nC_n|n\rangle$ is the initial quantum
state of the quantized probe field and the collective ground
states $|b^{(\sigma)}\rangle=|b_1,b_2,...,b_{N_{\sigma}}\rangle$.
To facilitate further discussion, we decompose the quantum state
of photons in terms of the basis of coherent states:
$|\phi_0=\sum_nC_n|n\rangle=\sum_jC'_j|\alpha^{(j)}\rangle$, thus
we have
\begin{eqnarray}\label{eqn:ds4}
|\Psi(0)\rangle=\sum_{j}C'_j|\alpha^{(j)}\rangle_{photon}\otimes|b^{(1)},
b^{(2)},...,b^{(m)}\rangle_{atom},
\end{eqnarray}
where $|\alpha^{(j)}\rangle=\sum_{n}P_n(\alpha^{(j)})|n\rangle$
with
$P_n(\alpha^{(j)})=\frac{(\alpha^{(j)})^n}{\sqrt{n!}}e^{-|\alpha^{(j)}|^2/2}$
is the probability of distribution function. After the quantum
memory process that the mixing angle $\theta$ is adiabatically
rotated from $0$ to $\pi/2$, the operator $\hat U_0(t)$ leads to
the resultant state
\begin{eqnarray}\label{eqn:ds5}
|\Psi(t)\rangle_0=|0\rangle_{photon}\otimes\sum_{j}C'_j|\alpha_1^{(j)},
\alpha_2^{(j)},...,\alpha_m^{(j)}\rangle_{coherence},
\end{eqnarray}
where $\alpha^{(j)}_k/\alpha^{(j)}_l=\sqrt{N_k}/\sqrt{N_l}$ and
$|\alpha^{(j)}|^2=|\alpha^{(j)}_1|^2+|\alpha^{(j)}_2|^2+...+|\alpha^{(j)}_m|^2$.
One can straightly verify that on Fock-state basis of the total
atomic system, the above result can be rewritten as
$|\Psi(t)\rangle_0=|0\rangle_{photon}\otimes\sum_{j}C'_j|\alpha^{(j)}\rangle_{coherence}=
|0\rangle_{photon}\otimes\sum_{n}C_n|n\rangle_{coherence}$, which
means that the quantum states of the atom coherence are the same
with that of input probe light. However, the actual evolution is
governed by $|\psi(t)\rangle=\hat U(t)|\psi(0)\rangle$. Thus the
actual final state is
\begin{eqnarray}\label{eqn:ds6}
|\Psi(t)\rangle=|0\rangle_{photon}\otimes\sum_{j}C'_j|\bar{\alpha}_1^{(j)},
\bar{\alpha}_2^{(j)},...,\bar{\alpha}_m^{(j)}\rangle_{coherence},
\end{eqnarray}
where
$\bar{\alpha}^{(j)}_k/\bar{\alpha}^{(j)}_l=(g_k\sqrt{N_k}/g_l\sqrt{N_l})\lim_{\Omega_k,
\Omega_l\rightarrow0}\Omega_k(t)/\Omega_l(t)$ and
$|\alpha^{(j)}|^2=|\bar{\alpha}^{(j)}_1|^2+|\bar{\alpha}^{(j)}_2|^2+...+
|\bar{\alpha}^{(j)}_m|^2$.

It is easy to see that if $g_1=g_2=...=g_m$ and
$\Omega_1=\Omega_2=...=\Omega_m$, one has
$\bar\alpha^{(j)}_k=\alpha^{(j)}_k$ and
$|\langle\Psi(t)|\Psi_0(t)\rangle|^2=1$. But for the inhomogeneous
coupling case, one can verify that generally
$|\langle\Psi(t)|\Psi_0(t)\rangle|^2<1$, i.e. for present case
\begin{eqnarray}\label{eqn:leak2}
\xi=1-|\langle\Psi(t)|\Psi_0(t)\rangle|^2>0.
\end{eqnarray}

One can give an intuitive understanding of above results. Since
our model should be treated as multi-ensemble atomic system, the
quantum information is stored in many atomic ensembles when the
control field is turned off. Because the couplings of different
atomic ensembles are different, the quantum information of photons
cannot be stored ``homogeneously" in different atomic ensembles,
i.e. it is divided into many inhomogeneous parts in the atoms. In
other words, a non-symmetric atomic state is prepared after the
photons are stored. This is why there is infidelity during the
storage process. However, the result $\xi>0$ does not mean that
there is leakage of the coherence in the quantum memory process.
In fact, when the control field is adiabatically turned on again,
i.e. the mixing angle $\theta$ is rotated adiabatically from
$\pi/2$ to $0$, with the dark-state evolution we find the quantum
states of the photons can be recovered from the atom coherence
\begin{eqnarray}\label{eqn:recover}
&|\Psi(t)\rangle\rightarrow|\Psi(0)\rangle:\nonumber\\
&|0\rangle_{photon}\otimes\sum_{j}C'_j|\bar{\alpha}_1^{(j)},
\bar{\alpha}_2^{(j)},...,\bar{\alpha}_m^{(j)}\rangle_{coherence}\longrightarrow\\
&\longrightarrow\sum_{n}C_n|n\rangle_{photon}\otimes|b^{(1)},
b^{(2)},...,b^{(m)}\rangle_{atom},\nonumber
\end{eqnarray}
The above derivation clearly shows that the quantum memory scheme
is still reversible in the inhomogeneous case. Even the storage
states of the atom coherence is different form the that of the
input quantized probe light due to the inhomogeneous coupling, the
quantum states of the photons can be recovered in the released
process when the mixing angle $\theta$ is rotated back to $0$ by
turning on the control field. This means that there really is
advantage of using collective spin states for quantum memory and
quantum information processing with EIT technique.

We make a few remarks on above results: from the eq. (\ref{eqn:1})
one can see that the inhomogeneous coupling leads to asymmetry in
the interaction Hamiltonian, i.e. $\hat H$ is not symmetric with
respect to permutation $\hat A_j\leftrightarrow\hat A_k$ (or $\hat
T_j\leftrightarrow\hat T_k$) of any two collective atomic
operators, thus the atomic state provided by this Hamiltonian is
also nonsymmetric. Since the initial state of the probe photons is
generally symmetric, the asymmetry of $\hat H$ will lead to
``leakage" of the quantum information when the photons are stored
into the atomic coherence. This is in agreement with the two-level
case \cite {sun}. However, as we have shown above, the dark state
for our EIT model exists even in the inhomogeneous coupling case,
thus the total state can evolve back into the initial one and the
quantum states of the photons can fully be recovered. Because
there is no dark state in two-level system, this phenomenon does
not occur in Sun $et al$ case and the leakage of quantum
information is hard to cancel out. The perfect character of EIT
quantum memory scheme with inhomogeneous coupling also accounts
for the existence of decoherence-free subspace (DFS)
\cite{Lidar,DFS1,DFS2,DFS3,DFS4} (i.e. here it is the DSS) in this
case. Finally, since the nonsymmetric entangled atomic states may
be observed with no decoherence induced by the inhomogeneous
coupling \cite{kuzmich}, the issue to measure the quantum states
of the atom coherence with nonsymmetric observables after the
photons are stored for current EIT model will be interesting and
deserve further study.

Before conclusion we should emphasize that here we consider the
quasi-inhomogeneous coupling case, which of course, is reasonable
for many practical cases. However, for the most general case that
every atom has a different coupling constant from others, there
may really be docoherence in the quantum memory process with EIT.
This is also an interesting issue that will be studied in our
future publications.

\section{conclusions}

We have discussed in detail the effect of inhomogeneous coupling
in a three-level EIT quantum memory process. The current model is
shown to be equivalent to a many-atomic-ensemble (sub-ensembles)
one, for which a DSS can exist even if the atomic state is far
from being a symmetric one. Although the inhomogeneity can lead to
a non-zero leakage during the storage process for the quantum
information of probe light, the quantum states can be fully
recovered in the released process due to the existence of present
non-symmetric DSS. This means that the EIT quantum memory
technique functions perfectly even for the inhomogeneous coupling
case. Furthermore, the model with sub-ensembles of atoms having
the same coupling to the field is powerful and in principle able
to deal with propagational and light scattering effects that are
often ignored (see, e.g. \cite{madsen}). Finally, based on our
sub-ensemble model we shall be able to propose a general algebraic
method to study all kinds of cases in the EIT technique, which
will be useful for probing the applications to quantum memory and
quantum information processing.

\bigskip

\noindent We thank the Prof. A.C. Doherty and Hui Jing for their
valuable discussions. We also thank the Prof. C. H. Oh for his
careful reading of the English of this paper. This work is
supported by NSF of China under grants No.10275036, and by NUS
academic research Grant No. WBS: R-144-000-071-305.
%\section{}

%\section{Results}
%\section{Conclusions}
%\indent
%ÕýÎÄ

%%%%%%%%%%%%%%%%%%%%%%%%%%%%%%%%%%%%%%%%%%%%%

%%%%%%%%%%%%%%%%%%%%%%%%%%%%%%%%%%%%%%%%%%%%%

\bigskip

\noindent

\end{document}